\documentclass[twocolumn,nofootinbib,showpacs,preprintnumbers,superscriptaddress] {revtex4}

\usepackage{amssymb,amsmath,amsfonts,amsbsy,graphicx}
\usepackage{color}
\usepackage{enumerate}
\usepackage{ulem}

\usepackage{enumerate}
\usepackage{slashed}
\usepackage{braket}

\newcommand{\dis}[1]{\begin{equation}\begin{split}#1\end{split}\end{equation}}
\newcommand{\be}{\begin{equation}}
\newcommand{\ee}{\end{equation}}
\def\bea{\begin{eqnarray}}
\def\eea{\end{eqnarray}}
\newcommand\ba{\begin{eqnarray}}
\newcommand\ea{\end{eqnarray}}

\newcommand{\eq}[1]{Eq.~(\ref{#1})}

\newcommand{\bfrac}[2]{{\left(\frac{#1}{#2} \right)  }}

\newcommand\tev{\,{\rm TeV}}
\newcommand\gev{\,{\rm GeV}}
\newcommand\mev{\,{\rm MeV}}

\newcommand{\mpl}{m_{\rm Pl}}

\newcommand\axino{{\tilde{a}}}
\newcommand\maxino{{m_{\axino}}}
\newcommand\neutral{\tilde{\chi}_1^0}
\newcommand\mneutral{m_{\tilde{\chi}_1^0}}

\begin{document}

\title{Searching for Axino-Like Particle at Fixed Target Experiments}

\author{Ki-Young Choi}
\email{kiyoungchoi@skku.edu}
\affiliation{Department of Physics, Sungkyunkwan University,  2066, Seobu-ro, Jangan-gu, Suwon-si, Gyeong Gi-do, 16419 Korea}

\author{Takeo Inami}
\email{inami@phys.chuo-u.ac.jp}
\affiliation{Department of Physics, Sungkyunkwan University,  2066, Seobu-ro, Jangan-gu, Suwon-si, Gyeong Gi-do, 16419 Korea}
\affiliation{Theoretical Research Division, Nishina Center, RIKEN, Wako 351-0198, Japan}
\affiliation{Institute of Physics, Vietnam Academy of Science and Technology, 
10 Dao Tan, Ba Dinh, Hanoi}

\author{Kenji Kadota}
\email{kadota@ibs.re.kr}
\affiliation{Center for Theoretical Physics of the Universe, Institute for Basic Science (IBS), Daejeon, 34051, Korea}

\author{Inwoo Park}
\email{inwpark@kaist.ac.kr}
\affiliation{Department of Physics, Korea Advanced Institute of Science and Technology, Daejeon, 34141,Republic of Korea}

\author{Osamu Seto}
\email{seto@particle.sci.hokudai.ac.jp}
\affiliation{Institute for the Advancement of Higher Education, Hokkaido University, Sapporo 060-0817, Japan}
\affiliation{Department of Physics, Hokkaido University, Sapporo 060-0810, Japan}

\begin{abstract}
We investigate the detectability of axino-like particle, which is defined as a supersymmetric partner of axion-like particle and can be a good candidate for dark matter in our Universe. Especially, we consider the fixed target experiments to search for the light axino-like particle with a neutralino as the next-to-lightest supersymmetric particle. We calculate the production and decay rate of neutralinos and the consequent number of events (such as photons and charged leptons) that are produced when the neutralinos decay to the axino-like particles.
\end{abstract}
 

\preprint{EPHOU-19-003, CTPU-PTC-19-08}
\maketitle
 

\section{Introduction}
\label{intro}

Supersymmetry (SUSY) gives a natural candidate for dark matter since the lightest supersymmetric particle (LSP) is stable when the R-parity is unbroken. When the axion is introduced to solve the strong CP problem in the standard model (SM), the fermionic SUSY partner, axino, can be the LSP and cold dark matter~\cite{Covi:1999ty,Covi:2001nw,Choi:2011yf,Bae:2011jb}. In the beyond standard models or in the superstring motivated models, there can exist ubiquitous axion-like particles (ALPs) and their fermionic superpartner, axino-like particle (which we call ``ALPinos''), with a wide range of mass scales~\cite{Witten:1984dg,Conlon:2006tq,Svrcek:2006yi,Arvanitaki:2009fg,Arias:2012az,Bae:2019vyh}. 

One of the stringent constraints on the interaction of axino comes from the lower bound on the Peccei-Quinn scale, $f_a \gtrsim 10^9\gev$~\cite{Kim:2008hd}. This constraint is obtained from the energy loss of stars due to the missing energy through the light axions. It is dominant for the mass of ALPs smaller than sub-MeV, and the  constraints for the mass range between sub-MeV and sub-GeV are given by SN1987A, supernovae and various beam-dump experiments depending on the size of $f_a$ and couplings to other particles, see e.g., Refs~\cite{Giannotti:2010ty,Jaeckel:2017tud}. For a recent comprehensive review on the constraints for the ALPs, refer to Refs.~\cite{Bauer:2017ris,Dolan:2017osp,Beacham:2019nyx} and for dark axion to Ref.~\cite{deNiverville:2019xsx}.
For ALPs~\cite{Rubakov:1997vp,Dienes:1999gw} heavier than about $1$ GeV, those bounds mentioned above are not applied and much smaller Peccei-Quinn scale can be allowed. Therefore, the interaction of ALPinos also can be free from this constraint as long as ALPs mass is larger than about $1$ GeV as we are going to assume.

In this paper, we study the possibility to detect the ALPinos from the search of hidden particles  in the fixed target experiments as well as the cosmological observations. Here, we focus on the ALPino ($\axino$)-photon ($A_{\mu}$)-photino ($\tilde{\gamma}$) interaction~\cite{Covi:2001nw}, of
\dis{
\mathcal{L_{\rm int}} = \frac{\alpha_{\rm em} C_{a \gamma\gamma}}{16 \pi f_a} \axino \gamma_5[\gamma^\mu, \gamma^\nu] \tilde{\gamma} F_{\mu\nu},
\label{Lint}
}
where $\alpha_{\rm em}$ is the fine-structure constant, $F_{\mu\nu}$ is the field strength of electromagnetic vector potential $A_{\mu}$ and $C_{a \gamma\gamma}= \mathcal{O}(1)$ is a model-dependent constant~\footnote{Compared to the usual notation $g_{a\gamma\gamma} a F_{\mu\nu}\tilde{F^{\mu\nu}}$, $g_{a\gamma\gamma} = \alpha_{\rm em}C_{a \gamma\gamma}/(8\pi f_a) $ in our model.}.
For the bino-like neutralino $\tilde{\chi}$, that will be studied in this paper, the coupling is multiplied by $\cos\theta_W$.
The interactions for ALPinos are typically very weak, suppressed by the scale of the spontaneous symmetry breaking $f_a$.
The mass of ALPino $m_{\tilde{a}}$ in general does not have to be of the order of the SUSY breaking, unlike the gravitino and ordinary supersymmetric particles, and $m_{\tilde{a}}$ can be indeed much smaller than the SUSY breaking scale in the visible sector~\cite{Chun:1992zk,Chun:1995hc,covi2001,Kim:2012bb}. While the model details such as the exact form of superpotential and the SUSY breaking mechanism are required to specify the value of $m_{\tilde{a}}$, in this paper, we consider the parameter range of $m_{\tilde{a}}$ ($\lesssim {\cal O}(1)$ GeV) which is small enough to be produced in the fixed target experiments.

We consider the case that the lightest neutralino $\neutral$ is the next-to-LSP (NLSP) and produced by decays of mesons at the target of the fixed target experiments. The neutralino goes through the shielding and  subsequently decays into an ALPino and a photon or a pair of charged particles, that can be probed at the detector far behind the target. We first consider the constraints from the present experiments such as CHARM~\cite{Bergsma:1983rt, Gninenko:2012eq}, NOMAD~\cite{Astier:2001ck, Gninenko:2011uv}.  Next, as a specific future prospect, we consider NA62~\cite{Dobrich:2018ezn},  SeaQuest~\cite{Berlin:2018pwi} and the SHiP experiment~\cite{Alekhin:2015byh,Anelli:2015pba} to search for ALPinos.
We also examine the relic density of ALPino as the main dark matter components in the Universe.

In Section~\ref{production}, we introduce the production of neutralinos in the fixed target experiments.
In Section~\ref{ALPinoFT}, we study the constraints from the existing experiments and detectability of ALPinos in the future experiments.
In Section~\ref{cosmology} we examine the relic density of ALPinos as dark matter. We conclude in Section~\ref{conclusion}.

\section{Production of neutralinos from  decay of mesons at fixed target experiment}
\label{production}

In the fixed target experiments, energetic particles of high-luminosity collide with the target and produce a large number of hadrons and weakly interacting particles. The weakly interacting particles can penetrate the shield and could give observable signatures at the far detector.
For example, a future fixed target experiment, Search for Hidden Particles (SHiP) experiment,  will use the CERN SPS beam with the energy $400$ GeV, which corresponds to the center of mass energy $\sqrt{s} \simeq 27.4 \gev$.
The total number of protons on the target is expected to be $N_{\rm pot}=2\times 10^{20}$~\cite{Alekhin:2015byh,Anelli:2015pba}. From the collision at the target, neutralinos with the mass less than ${\mathcal O}$(10) GeV can be  produced, and penetrate the shield. If these particles decay before the detector, which is about 100 meters away from the target, we can identify their decay. The characteristics of other similar experiments are summarized in Table~\ref{Table_Exp}.

The main production of neutralinos is a decay of mesons $M$ produced by collisions between the beam and the target, 
\dis{
pp \rightarrow M + X \quad  \textrm{with} \quad  M\rightarrow 	\neutral \neutral+X', 
}
with accompanying by-products $X , X'$ that are stopped at the shielding. 
The neutralinos penetrate the shield and decay in the decay-volume into an ALPino and a photon (or a pair of charged leptons):
\dis{
\quad \neutral \rightarrow \left\{
\begin{array}{ll}
 \axino + \gamma \\
 \axino+  \ell^+ + \ell^- \\
\end{array} 
\right. .
}
These photons or di-lepton can be detected at the far detector.
On the other hand, ALPinos can be produced also directly from off-shell photons, $\gamma^*\rightarrow \tilde{a} + \tilde{B}$, emitted from the beam in bremsstrahlung.
The cross section for electron beams is estimated as
\dis{
\sigma_\mathrm{Brem} \sim \left(\frac{\alpha_\mathrm{em}^3}{16 \pi f_a}\right)^2 \sim 10^{-10}\left(\frac{200 \mathrm{GeV}}{f_a}\right)^2 \mathrm{pb} .
}
Thus, existing electron beam experiments such as the E141~\cite{Bjorken:1988as} do not give significant constraints on the parameter region of our interest.
Similarly, the proton-nucleon cross section with off-shell bremsstrahlung photon can be expressed as
\dis{
\sigma_\mathrm{Brem} \sim \sigma_{pN}\alpha_\mathrm{em} \left(\frac{\alpha_\mathrm{em} E}{16 \pi f_a}\right)^2 \sim 10^{-15}\sigma_{pN} \left(\frac{200 \mathrm{GeV}}{f_a}\right)^2
\label{brem},
}
 with $\sigma_{pN}$ and $E$ are the total cross section for a proton-nucleon collision and the energy scale of interactions respectively.
As we will show, this contribution is sub-dominant compared with the production from meson decay.
Thus, we neglect this contribution in this paper.

\begin{table}
\begin{center}
\begin{tabular}{|c|c|c|c|c|}
\hline
Eperiment & Beam Eenergy& $N_\mathrm{pot}$ & $l$(m) & $\Delta l$ (m)   \\
\hline
CHARM~\cite{Bergsma:1983rt, Gninenko:2012eq}&400 GeV&$2.4 \times 10^{18}$&100&10\\
\hline
NOMAD~\cite{Astier:2001ck, Gninenko:2011uv}&450 GeV&$4.1 \times 10^{19}$&835&10\\
\hline
NA62~\cite{Dobrich:2018ezn} & 400 GeV & $10^{18}$ & 81 & 135  \\
\hline
SeaQuest~\cite{Berlin:2018pwi} & 120 GeV & 1.44 $\times$ $10^{18}$ & 5 & 10  \\
\hline
SHiP~\cite{Alekhin:2015byh,Anelli:2015pba} & 400 GeV &$2\times 10^{20}$&70&55\\
\hline
\end{tabular}
 \caption{The characteristics of the fixed target experiments used in this paper.}
\label{Table_Exp}
\end{center}
\end{table}

\begin{table}
\begin{center}
\begin{tabular}{ |c|c|c|c|c|c|c|c|c|}
\hline
  Meson & $\pi^+$ & $\pi^0$ & $\pi^-$  & $ \eta $ & $ \rho^0 $ &  $ \omega $ & $ \phi $\\ 
\hline 
$N_{M,multi}$ & $4.10$ & $3.87$ & $3.34$ & $0.30$  & $0.385$  & $0.390$ & $0.019$ \\ 
 \hline
\end{tabular}
\caption{The multiplicities  for mesons of $\pi, \eta, \rho, \omega,$ and $\phi$ in a proton-proton collision at $\sqrt{s} = 27.4 \gev$~\cite{Becattini:1997rv}.
\label{Table_multi}}
\end{center}
\end{table}

We estimate the number of produced mesons at the target by using their multiplicities $N_{M,multi}$  
 \dis{
N_M = N_{\rm pot} \times N_{M,multi}.
}
In Table~\ref{Table_multi}, we show the multiplicities for mesons of $\pi, K, \eta, \rho, \omega,$ and $\phi$ in a proton-proton collision at $\sqrt{s} \simeq 27.4 \gev$. For mesons, we also used the energy spectrum from a proton beam dump experiment~\cite{Bergsma:1985qz}.  

For the $J/\psi$ meson and the neutral $B$ meson, we estimate the number of produced mesons using the production cross section as~\cite{Alekhin:2015byh}
\dis{
N_M = N_{\rm pot} \frac{\sigma_M }{\sigma_{pN}},
}
where $\sigma_M$ is the meson production cross section per nucleon.
For $\sigma_{pN} \simeq 40$ mb,  $\sigma_{J/\psi}= 200$ nb and $\sigma_B=3.6$ nb, we find that  $10^{15}$ $J/\psi$ mesons  and $2\times 10^{13}$ neutral $B$-mesons are expected to be produced~\cite{Alekhin:2015byh}.

The number of produced neutralinos from a meson decay can be given by 
\dis{
N_{\neutral} \simeq 2N_M \times {\rm BR} (M\rightarrow \neutral \neutral +X'),
}
where the factor $2$ arises because a pair of neutralinos are produced from one meson decay.

The branching ratio of the decay for neutral mesons 
 have been calculated in the R-parity conserving case in Refs.~\cite{Borissov:2000eu,Dreiner:2009er}. 
Those are given by
\dis{
  & \mathrm{BR}(P \rightarrow \neutral \neutral)=C_P\left(\frac{1{\tev}}{m_{\tilde{q}}}\right)^4\left(\frac{m_{\neutral}}{1\mev}\right)^2
 \sqrt{1-4\frac{m_{\neutral}^2}{m_P^2}},\\
  & \mathrm{BR}(V \rightarrow \neutral \neutral)=C_V\left(\frac{1\tev}{m_{\tilde{q}}}\right)^4\left(1-4\frac{m_{\neutral}^2}{m_V^2}\right)^{3/2},
 \label{Eq_BR} }
 with $C_P=(9.30,0.263)\times 10^{-18}$ for pseudo-scalar meson $P=\pi^0, \eta$, and $C_V=(0.801, 15.7, 7.51,5.12\times 10^5, 4.47\times 10^6)\times10^{-18}$  for vector-meson $V=\rho^0, \omega, \phi, J/\psi, \Upsilon$, respectively.
By comparing Eqs.~(\ref{brem}) and (\ref{Eq_BR}), for example for $100$ MeV neutralinos, it becomes clear that the number of neutralinos produced bremsstrahlung is smaller than that from meson decays.
 
The analytical formulae to treat the decays 
$B^- \rightarrow K^-/\pi^-  \neutral \neutral$ was not available in the literature while the numerical results for the particular sets of SUSY particle masses have been presented in Refs.~\cite{Buras:2003jf,Jager:2008fc,Dreiner:2009er}.
We hence re-performed the calculations for our desirable parameter sets with much higher mass scale $\gtrsim {\cal O}(10)$ TeV than those (of the order of $100$ GeV) in the previous literature due to the tight constraints from the LHC.

Among those mesons, 
let us here outline the calculations we performed to treat the decays of a $B^-$ into neutralinos under the mass spectrum of supersymmetric particles which is consistent with the latest LHC results.
For concreteness, we assume that the lightest neutralino of bino-like is light enough to be produced at a fixed target experiment
 by taking the mass $M_1$ about $100$ MeV, while the other neutralino species are heavy.
We also take the masses of squark as $3 \tev$ and the other SUSY particles as $10$ TeV.

\begin{table}
\begin{center}
\begin{tabular}{| c| c| c|}
\hline
 Meson Decay& $B^-\rightarrow K^-\tilde{\chi}^0_1\tilde{\chi}^0_1$ \\ 
 \hline \hline
 $|C_F|$  & $3.22\times10^{-7}$ \\  
 $|C_L|$  & $2.50\times10^{-9}$ \\   
 $|C_R|$  & $9.58\times10^{-9}$ \\
 \hline
\end{tabular}
 \caption{The numerical values for coefficients $C_F, C_L$, and $C_R$ for $m_{\tilde{q}_i}=3\tev$ and $10 \tev$ for the masses of  other SUSY particles. }
\label{Table_Coeff}
\end{center}
\end{table}

The amplitude for the decay ${P_i^- \rightarrow P_f^- \tilde{\chi}^0_1\tilde{\chi}^0_1}$ is given by~\cite{Dreiner:2009er}
\dis{&\mathcal{M}_{P_i^- \rightarrow P_f^- \tilde{\chi}^0_1\tilde{\chi}^0_1}=\frac{G_F}{2}\Bigg(C_F(\bar{\tilde{\chi}}_1^0\gamma_\mu\gamma^5\tilde{\chi}_1^0)\bra{P_f}\bar{q}_f\gamma^\mu q_i\ket{P_i}\\
& + C_L(\bar{\tilde{\chi}}_1^0P_L\tilde{\chi}_1^0)\bra{P_f}\bar{q}_fq_i\ket{P_i}+C_R(\bar{\tilde{\chi}}_1^0P_R\tilde{\chi}_1^0)\bra{P_f}\bar{q}_fq_i\ket{P_i}\Bigg)\\
&+\frac{C_{p_s}}{m_{P_i}}(\bar{\tilde{\chi}}_1^0\slashed{p}_s\gamma^5\tilde{\chi}_1^0)\bra{P_f}\bar{q}_fq_i\ket{P_i},
\label{M0} 
}
where $q_i$ ($q_f$) is the initial (final) quark field and $G_F$ is the Fermi constant and $P_i^-$ and $P_f^-$ represent the initial and the final pseudo-scalar mesons.  For instance, for $K^-\rightarrow\pi^-\tilde{\chi}^0_1\tilde{\chi}^0_1$, those are $q_i=s$ and $q_f=d$. 
Here we note that the terms of $\bar{\tilde{\chi}}_1^0\gamma^\mu\tilde{\chi}_1^0$ and $\bar{\tilde{\chi}}_1^0[\gamma^\mu,\gamma^\nu]\tilde{\chi}_1^0$ vanish for a Majorana neutralino, and the terms $\bar{q}_f\gamma^\mu\gamma_5q_i$ and $\bar{q}_f\gamma_5q_i$ also vanish in the matrix element for pseudo-scalar mesons due to the odd parity.

Then, the invariant matrix element for a charged meson decay can be rewritten by~\cite{Borissov:2000eu,Dreiner:2009er}
\dis{&\mathcal{M}_{P_i^- \rightarrow P_f^- \tilde{\chi}^0_1\tilde{\chi}^0_1}=G_F\Bigg[C_F(\bar{u}_{\tilde{\chi}_1^0}\gamma_\mu\gamma^5v_{\tilde{\chi}_1^0})F^\mu\\
&+C_L(\bar{u}_{\tilde{\chi}_1^0} P_L v_{\tilde{\chi}_1^0})\frac{F\cdot (k_{P_i}-k_{P_f})}{m_{q_i}+m_{q_f}}\\
&+C_R(\bar{u}_{\tilde{\chi}_1^0} P_R v_{\tilde{\chi}_1^0})\frac{F\cdot (k_{P_i}-k_{P_f})}{m_{q_i}+m_{q_f}}\\
&+C_{{p_s}} (\bar{u}_{\tilde{\chi}_1^0}  \slashed{p}_s \gamma^5 v_{\tilde{\chi}_1^0})\frac{F\cdot (k_{P_i}-k_{P_f})}{m_{q_i}+m_{q_f}} \Bigg],
\label{M} 
}
 where $F_\mu$ is a matrix element defined as
$F_\mu\equiv\langle P_f^-|\bar{q}_f\gamma_\mu q_i|P_i^-\rangle$.
For the decomposition of $F_\mu$, we refer to Ref.~\cite{Nam:2007fx} for Kaons and Ref.~\cite{Ball:2004ye} for B-mesons as well as the Appendix in Ref.~\cite{Dreiner:2009er}.
The second term is obtained from 
\be
\langle P_f^-|\bar{q}_f q_i|P_i^- \rangle=\frac{F\cdot (k_{P_i}-k_{P_f})}{m_{q_i}+m_{q_f}}.
\ee
by using  $k_{P_i}-k_{P_f}\sim k_{q_i}-k_{q_f}$.

In Table~\ref{Table_Coeff}, we show the numerical results for the absolute values of the coefficients $C_F, C_L$, and $C_R$ for $m_{\tilde{q}}=3\tev$ to avoid the LHC constrants and the other SUSY particle masses of $10\tev$. We do not show $C_{p_s}$ because it is smaller than the other terms.
Those coefficients can be obtained from the one-loop diagrams by integrating the heavy spectrum~\cite{Dreiner:2009er}.
For this, we have used \texttt{Feynarts}, \texttt{FormCalc} and \texttt{LoopTools}~\cite{Hahn:1998yk, Hahn:2006ig} with neglecting all external momenta which are negligible compared to SUSY scalar particle masses and the W boson mass.  
For the Kaon decay, the $C_F$ term is dominant,
while for the B-meson decay both $C_F$ and $C_R$ terms are important.
 
From \eq{M} and using the values in Tab.~\ref{Table_Coeff},
 we obtain the branching ratio for the decay 
\dis{
{\rm BR} (B^-\rightarrow K^-\tilde{\chi}^0_1\tilde{\chi}^0_1) \simeq  2.8\times 10^{-13},
}
for the small lightest neutralino mass $m_{\tilde{\chi}_1^0} \ll m_K$.
These branching ratios would change for large SUSY particle masses, inversely proportional to the quartic power of a squark mass because the squark propagator shows up at the tree level.

\begin{figure*}[!t]
\begin{center}
\begin{tabular}{cc} 
 \includegraphics[width=0.4\textwidth]{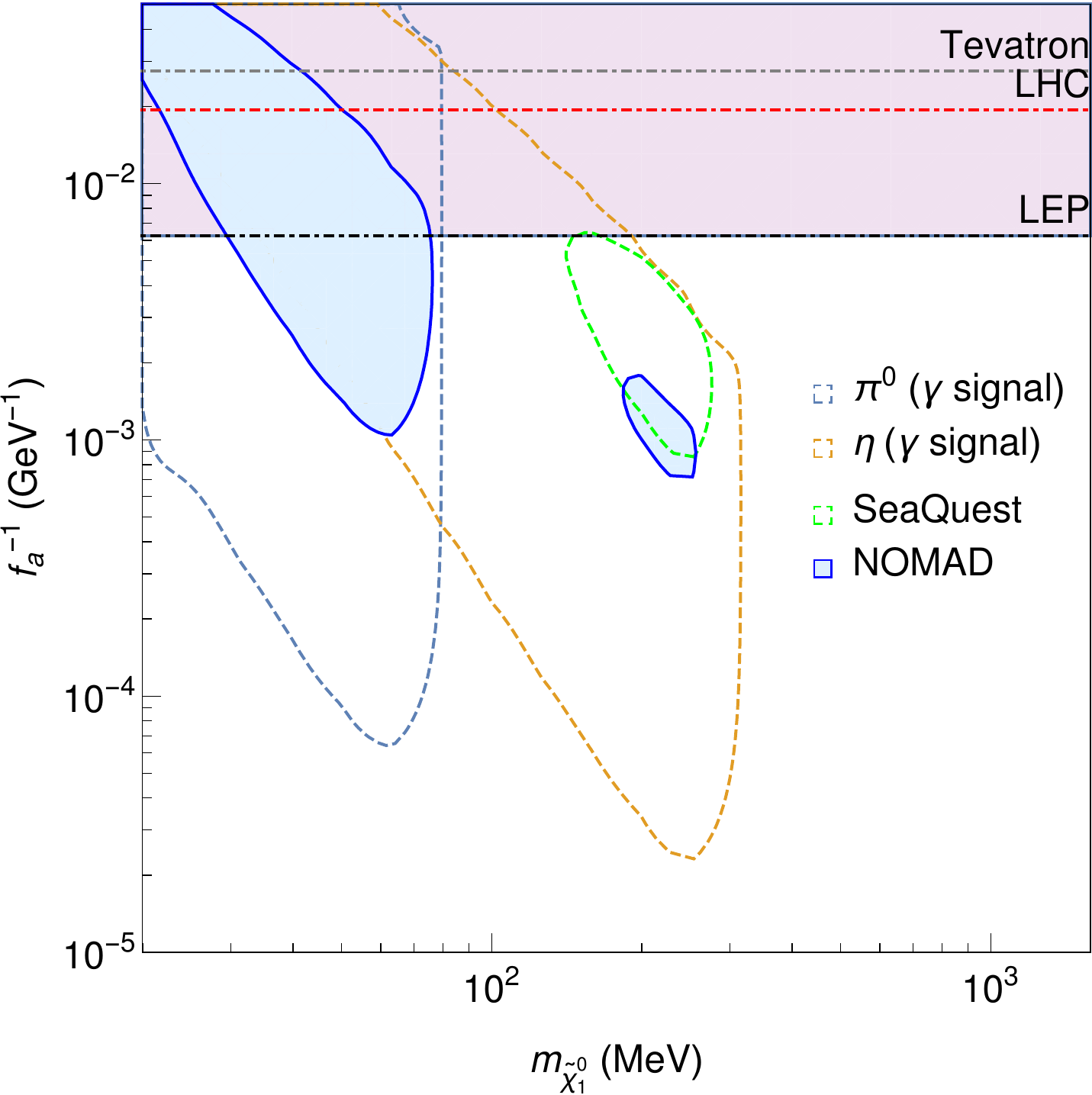}
 &
 \includegraphics[width=0.4\textwidth]{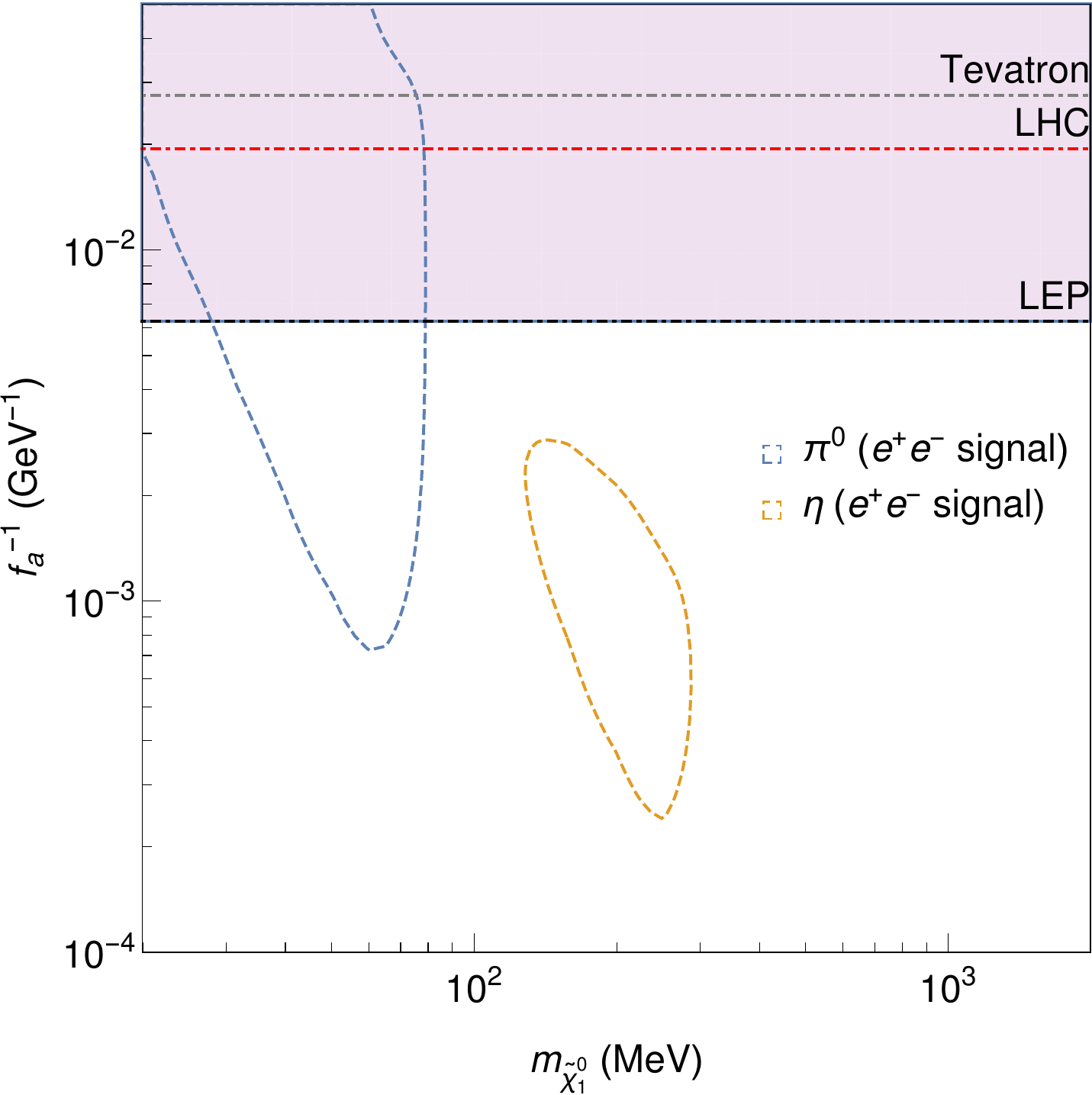}
   \end{tabular}
\end{center}
\caption{The region to be probed for searching  ALPino  in the plane of $f_a^{-1}$ vs neutralino mass $m_{\tilde{\chi}^0_1}$ using numbers of events of mono-photon (Left) and  electron-positron (Right). We took $\maxino=10 \mev$, $\tan{\beta}=10$ with masses of scalar quarks to be $3 \tev$ and those of other SUSY particles to be $10 \tev$.
The region above the horizontal lines are excluded by the corresponding collider experiments, such as Tevatron, LHC and LEP as indicated. The blue shaded region in the left panel can be constrained by the existing experiment of NOMAD (Blue) with assuming zero background of photons.  Inside the contour of dashed lines, the number of events can be larger than $3$ in the future experiments of  SeaQuest (Green dashed) and SHiP (blue and orange dashed). For SHiP, those are produced dominantly from the decay of $\pi^0$ (blue dashed) and $\eta$ (red dashed).
}
\label{ALPino}
\end{figure*}

\section{ALPino in the fixed target experiment}
\label{ALPinoFT}

The decay rate of the bino-like neutralino into an ALPino and a photon can be obtained from \eq{Lint} as
\begin{equation}
\Gamma(\neutral \rightarrow \axino +\gamma) = \frac{\alpha_{\rm em}^2 C_{a\chi\gamma}^2 }{128 \pi^3}\frac{\mneutral^3}{f_a^2} \left( 1- \frac{\maxino^2}{\mneutral^2} \right)^3 ,
\label{chigamma}
\end{equation}
with $C_{a\chi\gamma}= C_{a\gamma\gamma}Z_{\chi B}$
where $Z_{\chi B}$ stands for the Bino fraction of the neutralino.
In the following, we take the pure-Bino limit $Z_{\chi B} \rightarrow 1$, and 
 $C_{a\gamma\gamma}=1$ for concreteness.
In this case, the lifetime of the neutralino is given as
\dis{
\tau(\neutral \rightarrow \axino +\gamma) = &0.49 \times 10^{-9}{\rm sec}\, \bfrac{1/128}{\alpha_{\rm em}}^2\bfrac{f_a}{10^{5}\gev}^2\\
&\times \bfrac{10\gev}{\mneutral}^3 \left( 1- \frac{\maxino^2}{m_\chi^2} \right)^{-3} .
}
The neutralino can decay also into an ALPino and two charged leptons, $\neutral \rightarrow \axino + \ell^+ + \ell^-$, through a virtual photon and the Z-boson.  The decay rate for this is (refer to Appendix B for details)
\dis{
\Gamma(\tilde{\chi}_1^0 \rightarrow \axino + \ell^ + +  \ell^-)  \simeq \frac{\alpha_{\rm em}^3 C_{a\chi\gamma}^2 }{512 \pi^4}\frac{m_{\tilde{\chi}_1^0 }^3}{f_a^2}\left(4\ln{\frac{\mneutral}{m_\ell}}-6\right)\label{chiepm},
}
in the limit of $\mneutral\gg m_\ell$. Here $\ell$ refers to light charged leptons such as an electron and/or muon.

\begin{figure*}[!t]
\begin{center}
\begin{tabular}{cc} 
 \includegraphics[width=0.4\textwidth]{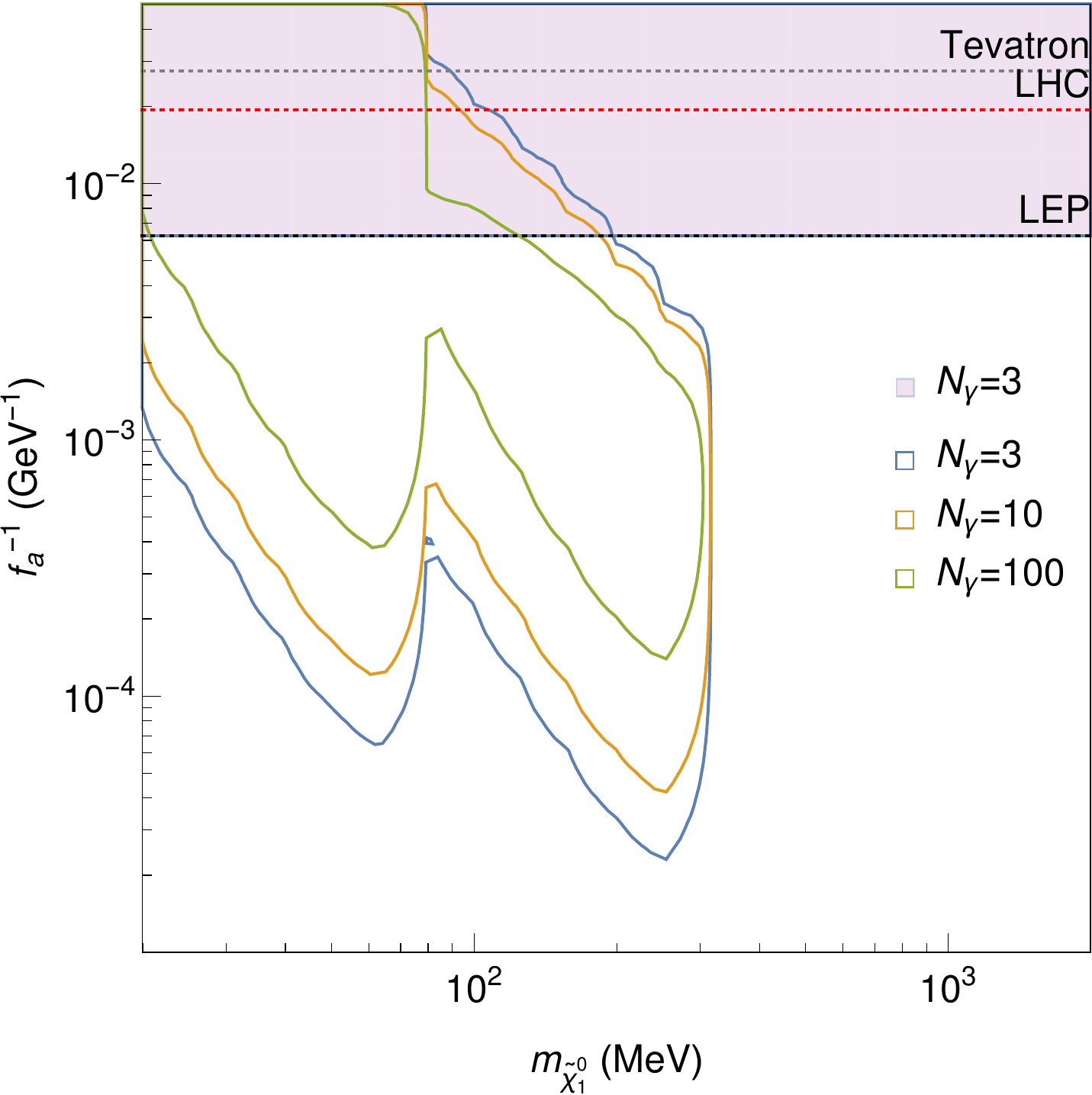}
 &
 \includegraphics[width=0.4\textwidth]{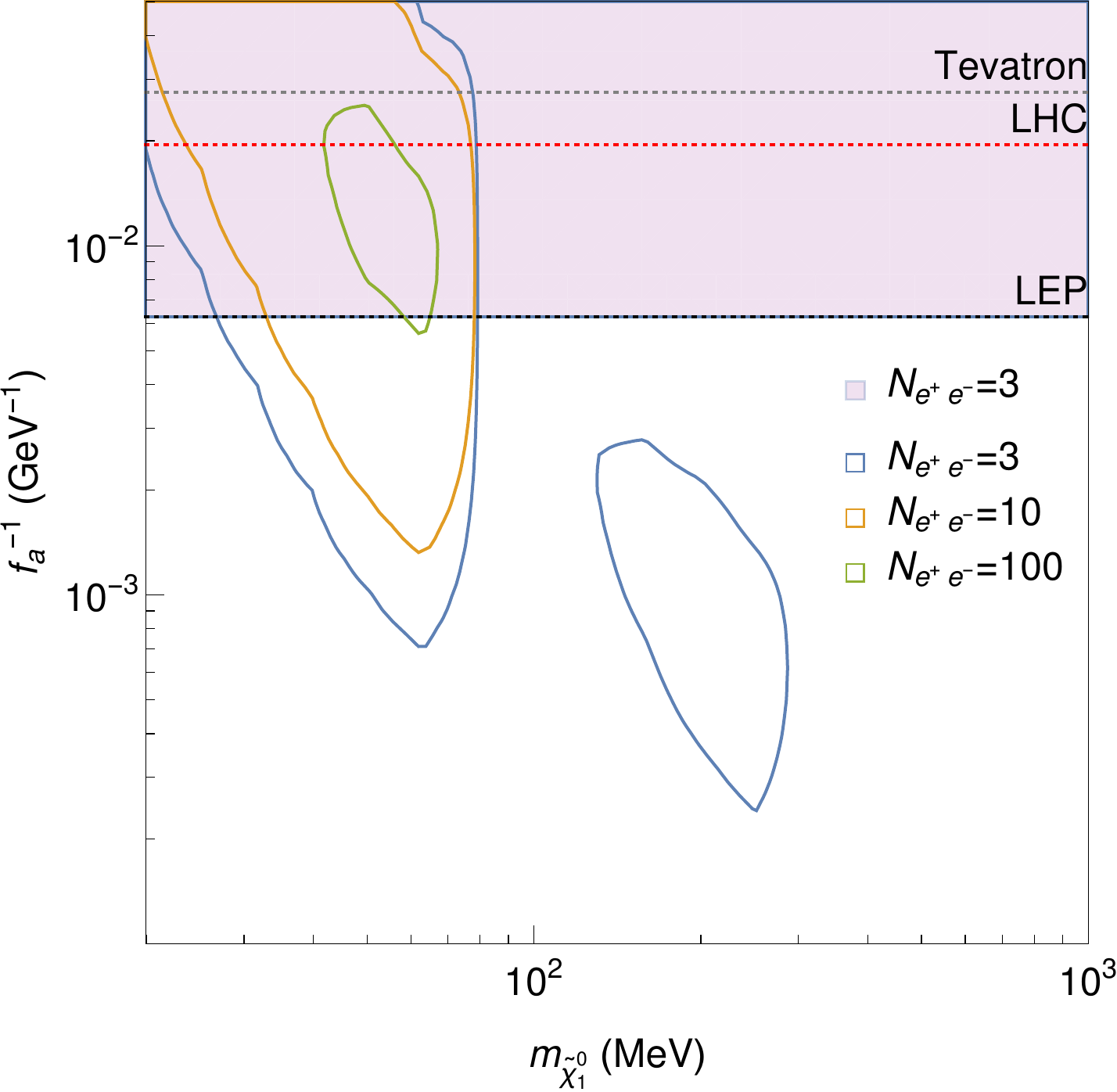}
  \end{tabular}
\end{center}
\caption{The number of signals in the future SHiP experiment with mono-photon (Left) and with  electron and positron (Right). We show the contour plots for the number of signals $3,10,100$ in each case with blue, red, and green color, respectively. We used the same mass spectrum as used in Fig.~\ref{ALPino}.}  
\label{ALPino_number}
\end{figure*}

The number of neutralinos that decay inside the detector region can be estimated
 by multiplying the decay probability with the number of neutralinos produced~\cite{Alekhin:2015byh}.
For this, we consider the energy distribution of the neutralinos from meson decays, the corresponding $\gamma$ factor and the angular acceptance $\tan\theta_c \lesssim d/(l + \Delta l)$  assuming that the parent meson has a momentum parallel to this axis,
 with the detector size $d$, 
 the decay volume with the distance $l$ from the target and its volume length of $\Delta l$.
For two body decays, the number of events in the detector region is estimated as
\dis{
N_{\rm det} &=N_{\tilde{\chi}^0_1}\int dE_M p_M(E_M)\int_0^{\theta_c} \frac12\sin\theta d\theta \int dE_\chi \, p(E_\chi,E_M)\\
&\times \left[ \exp\left( -\frac{l}{\gamma_\chi\beta_\chi c\tau} \right)-\exp\left( -\frac{l+\Delta l}{\gamma_\chi\beta_\chi c\tau} \right) \right].
}
 with $\gamma_\chi=E_{\tilde{\chi}_1^0}/\mneutral$ being the relativistic $\gamma$ factor and $\beta_\chi=v/c$.  
We integrate over the energy distribution of the neutralino $p_\chi(E_\chi,E_M)$ and also the energy spectrum of the meson $p_M(E_M)$. For two-body decay, the $p_\chi(E_\chi,E_M)$ is flat distribution in the range $\frac12(1-\beta\bar{\beta}) \leq E_\chi/E_M \leq \frac12(1+\beta\bar{\beta})$ with $\beta = \sqrt{1-M^2/E_M^2}$ and $\bar{\beta} = \sqrt{1-4m_\chi^2/M^2}$.
Here, we have restricted the angle less than $\theta_c \simeq \arctan(d/(l+\Delta l))$ which is the angle of the neutralino from the axis of the decay volume to the detector radius. For three-body decay,  the formula is more involved and summarized in the appendix.
We adopt the criteria $N_{\rm det} < 3$ for no observation, assuming the negligible background events.
However, for mono-photon this is a very optimistic assumption, since the background rejection of a single photon is quite difficult.
With related to this, in Fig.~\ref{ALPino_number}, we show also the contour of the number of events.
For mono-photon events, the background may be reduced by selecting the energy range and the angle acceptance.
In Fig.~1, we considered the event with the energy of the final charged particles or the photons larger than 1 GeV.

In Fig.~\ref{ALPino}, we show the results of our analysis, by the  region probed with mono-photon (left) and electron-positron signals (right). For the mono-photon signals, NOMAD, CHARM can constrain the region of neutralino mass less than 80 MeV and $f_a^{-1} \gtrsim 10^{-3}\gev^{-1}$ from pion decay. For NOMAD (inside the blue solid line), the small region at $m_{\tilde{\chi}^0_1}\sim 200 \mev$ also can be constrained due to its high luminosity. The experiments NuCal and BEBC also give the comparable constraints~  \cite{Blumlein:1990ay,CooperSarkar:1985nh}.

In the future experiments, we found NA62 does not improve much compared to NOMAD, so we did not show in the figure. However SeaQuest  (green dashed line) can probe a similar region  as NOMAD due to the  large angular acceptance with a near detector,
if ECAL is installed in the future upgrade of the detector and the photon background can be reduced with the proper adjustment of the shielding, which can also reduce the  fiducial volume of the detector though.
The  SHiP experiment is expected to have a better sensitivity.
The projected search region by the SHiP experiment is shown with dashed lines (blue for $\pi^0$ decay and orange for $\eta$ decay) for mono-photon signals (Left) and for lepton pair production signals  (Right) in Fig.~\ref{ALPino}, where the inside regions can be probed by the specified meson decay with the number of events more than $3$. With electron pairs, the decay of $\pi^0$ and $\eta$ may produce a detectable number of events for $f_a^{-1}\gtrsim 2\times 10^{-5} \mathrm{GeV}^{-1}$ and $m_{\tilde{\chi}_1^0}\lesssim 300 \mev$ from the mono-photon and $f_a^{-1}\gtrsim 2\times 10^{-4} \mathrm{GeV}^{-1}$ for  the electron pair. Here, we have used $\maxino=10 \mev$ and $\tan\beta = 10$ (this $\tan\beta$ value is chosen in our analysis which can lead to the desirable Higgs mass while being consistent with the other bounds such as $B_s \rightarrow \mu^+ \mu^-$).

\begin{figure}[!t]
\begin{center}
\begin{tabular}{c} 
 \includegraphics[width=0.4\textwidth]{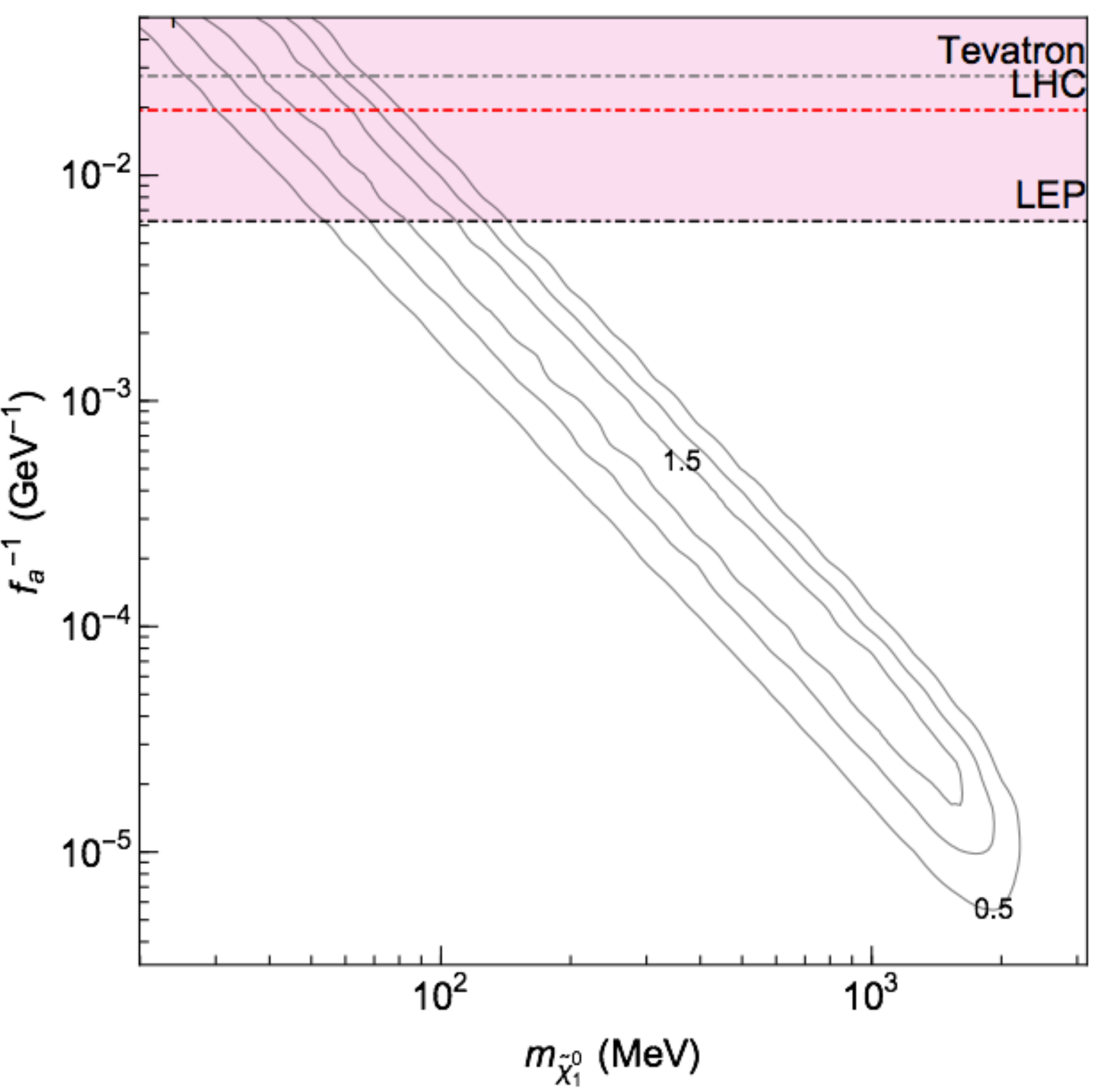}
   \end{tabular}
\end{center}
\caption{The number of photon event from the neutralinos produced by the decay of B-meson. The  number of events is smaller than 2 and may not be possible to be detected in the SHiP.}
\label{Bmeson}
\end{figure}

The ALPino decay constant is also bounded from other high-energy colliders. The bounds from the mono-photon (mono-jet) search from the lepton (hadron) colliders such as the TEVATRON, the LHC, and the LEP are shown with horizontal dotted lines in Fig.~\ref{ALPino}~\cite{fort2011}. The region above the horizontal line are disfavored~\footnote{Note that our axino coupling has the form of the dipole interaction
  \ba
  \frac{\alpha_{em}C_{a\gamma \gamma}}{16 \pi f_a}
  \bar{\tilde{a}} \gamma_5
  \left[
    \gamma^{\mu},
    \gamma^{\nu}
    \right]
  \tilde{\gamma}
  F_{\mu \nu}
  =
  -\frac{\alpha_{em}C_{a\gamma \gamma}}{8\pi f_a}
  \bar{\tilde{a}}
  \sigma_{\mu \nu}
  \tilde{\gamma}
  \tilde{F}^{\mu \nu}
  \ea
and Fig.~\ref{ALPino} shows the collider bounds on the dipole interactions discussed in Ref. \cite{fort2011} to which we refer the readers for the details.}.

The bounds on our ALPino LSP scenarios with the neutralino NLSP in this figure share the common features.  The largest value of the lightest  neutralino mass probed by the experiments (represented by the vertical edge in the figure) depends on the mass of the mesons which decays into the neutralino. The larger values of $f_a^{-1}$, which are not able to be probed in the figure, are due to a too short lifetime of the neutralino so that the ALPino production occurs before the neutralino reaching the detector. On the other hand, if $f_{a}^{-1}$ is too small, the neutralino decays after passing through the whole detector (represented by the lower end of $f_a^{-1}$ in the reach regions).
 The photon signals tend to give tighter bounds on $f_a^{-1}$ than $e^{\pm}$ signals because of the difference in the decay rates (Eqns.~(\ref{chigamma}) and~(\ref{chiepm})). The decay rate into $e^{\pm}$ is smaller than that into $\gamma$ because of the phase space suppression and additional vertex couplings.
We also note that the decay rate of a neutral meson ($M=\pi^0, \eta, \eta',$ $\rho^0, \omega, \phi, J/\psi, \Upsilon$) has a large dependence on the squark mass scale (inversely proportional to the quartic power of a squark mass because the squark propagator shows up at the tree level). 

In Fig.~\ref{ALPino_number}, we also show the number of signals in the $f_a^{-1}$ vs. $m_{\tilde{\chi}^0_1}$
 in the future SHiP experiment for mono-photon  (left window) and for electron and positron (right window) with the number of $3,10$, and $100$.
 
In Fig.~\ref{Bmeson}, we show the contour plot of the number of photons from neutralinos produced in B-meson decay observable at the SHiP experiment.
The number is not enough for a detection since the number is smaller than $2$ and thus it does not contribute in the Fig.~\ref{ALPino}. 
However, it may be interesting to check the detectability in the other future experiment such as MATHUSLA~\cite{Curtin:2018mvb,Alpigiani:2018fgd}, CODEX-b~\cite{Gligorov:2017nwh} and FASER~\cite{Ariga:2018zuc,Ariga:2018uku} or in the search for long-lived particles through displaced~\cite{Aaij:2017rft} or delayed~\cite{Liu:2018wte} tracks in the LHC experiments.

\section{Cosmology of ALPino}
\label{cosmology}
The stable ALPino can be a good candidate for dark matter, if they are  produced in the early Universe with a right amount.
With the coupling in \eq{Lint}, the freeze-out temperature of relativistic ALPino at high temperature is~\cite{Rajagopal:1990yx}
\dis{
T_f \simeq 1 \gev \, \bfrac{f_a}{10^{5}\gev}^2\bfrac{0.01}{\alpha_{\rm em}}^3,
\label{Tf}
}
where the ALPino and gaugino co-annihilation is the dominant interaction for the thermalization.

For $T_f > m_{\tilde{a}}$, the ALPino decouples when they are relativistic, and the abundance is determined by the effective degrees of freedom at that time.  
In this case, the relic number density of ALPino is similar to that of neutrinos as hot relic and 
the ALPinos end up being overabundant for the mass range of our interest around MeV to sub-GeV.
This conclusion holds in the case of $T_f < m_{\tilde{a}}$ as long as the abundance is determined by the freeze-out.
As is well known for the WIMP,
 the freeze-out occurs at around $T_f \simeq \mneutral/25$ for a weak scale masses of $\tilde{a}$ and $\tilde{\chi}$ and the weak scale cross section, 
 which would give the right amount of WIMP abundance.
However, neutralino-ALPino annihiation cross section $\sigma v \sim \alpha^3_\mathrm{em}/f_a^2$ is much smaller than
 about one picobarn. Thus, if neutralino and ALPino are freeze-out thermal relics, DM is overabundant.

\begin{figure}[t]
\includegraphics[width=0.4\textwidth]{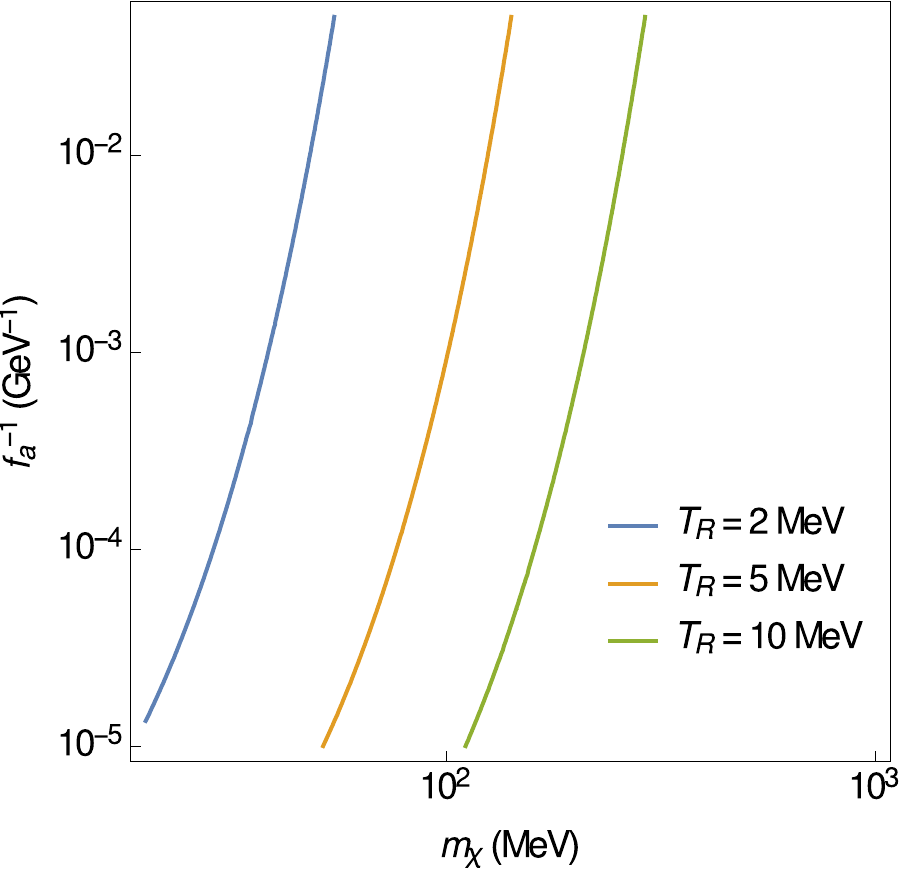}
\caption{The contours satisfying $\Omega_{\tilde{a}}h^2 \simeq 0.1$ are shown in the plane of $f_a^{-1}$ vs neutralino mass for two different cases of $T_R$. This plot is the case for $m_{\tilde{a}}=10$ MeV. By comparing with Fig.~\ref{ALPino}, we find that the parameter region to be probed by the SHiP experiment is cosmologically interesting from the viewpoint of dark matter. The lower bound on the reheating temperature, $T_R \gtrsim $ a few MeV, has been derived for successful Big Bang Nucleosysthesis~\cite{Kawasaki:1999na,Kawasaki:2000en,Ichikawa:2005vw,deSalas:2015glj,Hannestad:2004px,Ichikawa:2006vm,DeBernardis:2008zz,Hasegawa:2019jsa}
}
\label{Oh2}
\end{figure}

One possibility to obtain the correct dark matter abundance is
 the cosmology with a low reheating temperature, $T_R < T_f$, which is small enough so that
 neither neutralino nor ALPino can be thermalized~\cite{Roszkowski:2015psa}.
Nevertheless, ALPinos can be produced by scattering processes in thermal plasma.
The dominant production processes are $f \bar{f} \rightarrow \neutral \tilde{a}$
 through s-channel photon exchange
 followed by the neutralino decay $\neutral \rightarrow \tilde{a} \gamma$,
 where $f$ denote a SM fermion.
The number density of ALPino is evaluated by solving the following Boltzmann equation 
\dis{
 \frac{d n_{\tilde{a}}}{dt}+ 3 H n_{\tilde{a}} = C_\mathrm{coll.}(f \bar{f} \rightarrow \neutral \tilde{a}) +  C_\mathrm{decay}(\neutral \rightarrow \tilde{a}+\gamma) ,
}
 which can be rewritten as
\dis{
 Y_{\tilde{a}} = \int^{T_R}_{T_0} \frac{2 C_\mathrm{coll.}}{s H T} dT
}
 where $T_R$ is the reheating temperature, $s$ is the entropy density, $H$ is the Hubble parameter,
 respectively, and a factor $2$ comes from the R-parity conserving neutralino decay.
The resultant abundance depends on $m_{\tilde{a}}, \mneutral$ and $T_R$, and can be expressed in the leading order as
\dis{
\Omega_{\tilde{a}}h^2 \simeq & \, \frac{2.8 \times 10^{11}}{57 \pi^3}\left(\frac{90}{\pi^2 g_*}\right)^{3/2}
 \left(\frac{\alpha_\mathrm{em} }{8 \pi}\right)^3 \sum_f Q^2 \\ 
 & \times \left(\frac{m_{\tilde{a}}M_P}{f_a^2}\right)\frac{m_{\tilde{\chi}_1^0}}{\mathrm{GeV}} e^{-m_{\tilde{\chi}_1^0}/T_R} , 
}
 for $T_R \ll m_{\tilde{\chi}_1^0}$ with $g_*$ being the relativistic degrees of freedoms, $M_P$ being the reduced Plannck mass, and 
 $Q$ being the electromagnetic charge of initial state SM fermions in the unit of $e$.
The appropriate abundance $\Omega_{\tilde{a}}h^2 \simeq 0.1$ can be obtained
 for the mass range of our interest as shown in Fig.~\ref{Oh2}.
%
\section{Conclusion}
\label{conclusion}

In supersymmetric models with axion/ALP, there exists its fermionic superpartner, axino/ALPino.
For a heavy ALP, the usual astrophysical  is not applicable anymore and
 the symmetry breaking scale  known as  decay constant of ALP model, $f_a$, can be lower than the typical QCD-axion decay constant. 
In this paper, we have studied the possibility to search the light ALPino with the bino-like neutralino as the NLSP at fixed target experiments and also the possibility for the ALPino to be the dark matter. We found that the current experiments NA62 and SeaQuest will constrain  this model, and the future experiment SHiP can probe the ALPino decay constant as large as $5\times 10 ^3$ GeV from 
$e^+ e^-$ signals and $5\times 10^4$ GeV from mono-photon signals for the sub-GeV neutralino mass.
We note also that the ALPinos can be produced in the early Universe with the desirable amount of dark matter by the freeze-in from the thermal equilibrium at low-reheating temperature.

\appendix
\section{Three-body decay rate for  $\tilde{\chi}_1^0 \rightarrow \tilde{a}\ell^+\ell^-$.}
\begin{figure}
\begin{center}
\includegraphics[width=0.45\textwidth]{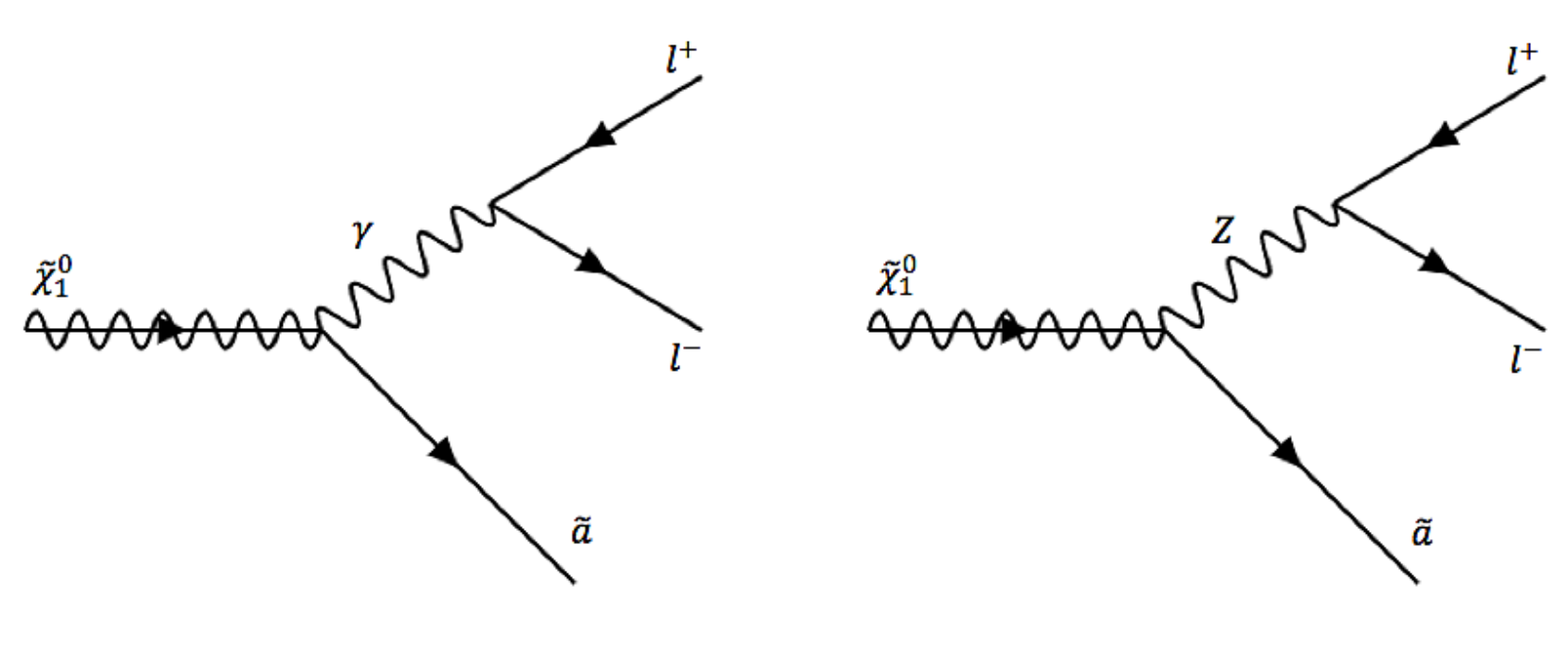}
\caption{The decay of neutralino to the axino and the lepton anti-lepton pair mediated by photon (Left) and Z boson (Right)
}
\label{fig:neudecay}
\end{center}
\end{figure}

The neutralino can decay into axino and  lepton, anti-lepton pair through photon and Z-boson, as shown in Fig.~\ref{fig:neudecay}.
Since the energy of the fixed target experiment is smaller than the mass of the Z-boson, the three-body is dominated by the photon-mediation and the Z-boson can be ignored.

\begin{widetext}
Then, the decay rate of $\tilde{\chi}_1^0 \rightarrow \tilde{a}\ell^+\ell^-$ is given by
\dis{\Gamma(\tilde{\chi}_1^0 &\rightarrow \tilde{a}+\ell^++\ell^-)=\\
\frac{\alpha_e^3C_{a\chi\gamma}}{512\pi^4}&\frac{1}{f_a^2m_{\tilde{\chi}_1^0}^3} \bigg[4(m_{\tilde{a}}^6-m_{\tilde{a}}^4 m_{\tilde{\chi}_1^0}^2-\maxino^2\mneutral^4+18\maxino\mneutral m_l^4-4m_l^6+\mneutral^6)\ln{\left(\frac{ \sqrt{(m_{ \tilde{a} }-m_{ \tilde{\chi}_1^0 })^2-4m_l^2}-m_{ \tilde{a} }+m_{ \tilde{\chi}_1^0 } }{ 2m_l}\right) }\\
&\frac{1}{3}\sqrt{\frac{(\maxino-\mneutral)^2-4m_l^2}{(\maxino-\mneutral)^2}}\bigg\{-18\maxino^6+21\maxino^5\mneutral+\maxino^4(30m_l^2+22\mneutral^2)-2\maxino^3(57m_l^2+25\mneutral^3)\\
&+\maxino^2(-12m_l^4+8m_l^2\mneutral^2+22\mneutral^4)+3\maxino(8m_l^4\mneutral-38m_l^2\mneutral^3+7\mneutral^5)-6(2m_l^4\mneutral^2-5m_l^2\mneutral^4+3\mneutral^6) \bigg\}\bigg].
}

For $\mneutral >> \maxino, m_l$, we have
\dis{\Gamma(\tilde{\chi}_1^0 &\rightarrow \tilde{a}+\ell^++\ell^-)=\frac{\alpha_{\rm em}^3 C_{a\chi\gamma}^2 }{512 \pi^4}\frac{m_\chi^3}{f_a^2}\left(4\ln{\frac{\mneutral}{m_l}}-6\right).
}

\section{Three-body decay}
The decay rate of three body decay is given by
\begin{equation}
\Gamma(M\rightarrow1+2+3) = \frac{1}{256\pi^3}\frac{1}{M^3}\int |\mathcal{M}(s_1,s_3)|^2 ds_{1} ds_{3},
\label{eq:decayrate}
\end{equation}
where invariant masses $s_i=(P-p_i)^2$. 
We can express the invariant masses in terms of the variables in the boosted frame by the meson velocity given by $E_M$ along the axis of the decay volume.

In this boosted frame, finally we have the number of neutralinos observable at the detector
\begin{align}
&N_{\rm det} = \frac{1}{128\pi^3} \frac{N_M}{M^2 \Gamma_{tot}} \int dE_Mp(E_M) \int_0^{\theta_c} \int_{E_3^{'-}(\theta;E_3^{max})}^{E_3^{'+}(\theta;E_3^{max})} \int_{s_{1-}(E_\chi,\theta)}^{s_{1+}(E_\chi,\theta)} \nonumber\\
&\qquad\times  |\mathcal{M}(s_1,E_\chi,\theta)|^2 \left[\exp\left(-\frac{l}{\gamma_\chi\beta_\chi c\tau}\right)-\exp\left(-\frac{l+\Delta l}{\gamma_\chi\beta_\chi c\tau}\right)\right] \sin\bar{\theta} ds_1 dE_\chi  d\theta,
\end{align}
where
\dis{
\gamma_\chi = E_\chi / m_\chi, \qquad  \beta_\chi = \frac{ \sqrt{E_\chi^2-m_\chi^2}}{ E_\chi}.
}
Here $\bar{\theta}$ and $\theta$ are the angles of the neutralino in the rest frame and boosted frame of the meson respectively. The integration ranges are given appropriately from the kinematics of three-body decay.

\section{ $f \bar{f} \rightarrow \tilde{\chi} \tilde{a}$ amplitude}

The spin averaged amplitude squared for $f \bar{f} \rightarrow \tilde{\chi} \tilde{a}$ via $s$-chennel photon mediation
 integrated over the Lorents invariant phase space of the final states is given by 
\begin{align}
 & \int \overline{|\mathcal{M}|^2} d\mathrm{LIPS} \nonumber \\
 = & \frac{e^2Q^2}{32\pi}\left(\frac{\alpha_\mathrm{em}}{8\pi f_a}\right)^2
 \frac{16}{3 s^3} \left( 
-2 m_{\tilde{\chi}} s^2 \left(6 m_{\tilde{a}}^3-7 m_{\tilde{a}}^2 m_{\tilde{\chi}}+2 m_{\tilde{\chi}}^3\right)
     + (m_{\tilde{\chi}}^4 s+s^3)(m_{\tilde{\chi}}-m_{\tilde{a}}) (m_{\tilde{\chi}}-5 m_{\tilde{a}})
+m_{\tilde{\chi}}^8+s^4 \right. \nonumber \\
& \left. + \left(m_{\tilde{\chi}}^2+s\right) \sqrt{4 m_{\tilde{a}}^2 s+\left(m_{\tilde{\chi}}^2-s\right)^2} \left(3 m_{\tilde{a}}^2 s+m_{\tilde{\chi}} s (m_{\tilde{\chi}}-6 m_{\tilde{a}})+m_{\tilde{\chi}}^4+s^2\right) \right),
\end{align}
 for $s \gg m_f^2$ , and 
\begin{align}
 \int \overline{|\mathcal{M}|^2} d\mathrm{LIPS} = \frac{e^2Q^2}{32\pi}\left(\frac{\alpha_\mathrm{em}}{8\pi f_a}\right)^2\frac{32}{3}(2 m_f^2+ s),
\end{align}
 for $s \gg m_{\tilde{\chi}}^2, m_{\tilde{a}}^2$, respectively.
\end{widetext}

\section*{Acknowledgment}
K.-Y.C. was supported by the National Research Foundation of Korea(NRF) grant funded by the Korea government(MEST) (NRF-2016R1A2B4012302). KK and I.Park were supported by Institute for Basic Science (IBS-R018-D1).  O.S. was in part supported by KAKENHI Grants No.~19K03860, No.~19H05091 and No.~19K03865. This project was supported by JSPS and NRF under the Japan - Korea Basic Scientific Cooperation Program (NRF-2018K2A9A2A08000127). 



\def\prp#1#2#3{Phys.\ Rep.\ {\bf #1} #2 (#3)}
\def\rmp#1#2#3{Rev. Mod. Phys.\ {\bf #1}  #2 (#3)}
\def\anrnp#1#2#3{Annu. Rev. Nucl. Part. Sci.\ {\bf #1} #2 (#3)}
\def\npb#1#2#3{Nucl.\ Phys.\ {\bf B#1}  #2 (#3)}
\def\plb#1#2#3{Phys.\ Lett.\ {\bf B#1}  #2 (#3)}
\def\prd#1#2#3{Phys.\ Rev.\ {\bf D#1}, #2  (#3)}
\def\prl#1#2#3{Phys.\ Rev.\ Lett.\ {\bf #1}  #2 (#3)}
\def\jhep#1#2#3{JHEP\ {\bf #1}  #2 (#3)}
\def\jcap#1#2#3{JCAP\ {\bf #1}  #2 (#3)}
\def\zp#1#2#3{Z.\ Phys.\ {\bf #1}  #2 (#3)}
\def\epjc#1#2#3{Euro. Phys. J.\ {\bf #1}  #2 (#3)}
\def\ijmp#1#2#3{Int.\ J.\ Mod.\ Phys.\ {\bf #1}  #2 (#3)}
\def\mpl#1#2#3{Mod.\ Phys.\ Lett.\ {\bf #1}  #2 (#3)}
\def\apj#1#2#3{Astrophys.\ J.\ {\bf #1}  #2 (#3)}
\def\nat#1#2#3{Nature\ {\bf #1}  #2 (#3)}
\def\sjnp#1#2#3{Sov.\ J.\ Nucl.\ Phys.\ {\bf #1}  #2 (#3)}
\def\apj#1#2#3{Astrophys.\ J.\ {\bf #1}  #2 (#3)}
\def\ijmp#1#2#3{Int.\ J.\ Mod.\ Phys.\ {\bf #1}  #2 (#3)}
\def\apph#1#2#3{Astropart.\ Phys.\ {\bf B#1}, #2 (#3) }
\def\mnras#1#2#3{Mon.\ Not.\ R.\ Astron.\ Soc.\ {\bf #1}  #2 (#3)}
\def\nat#1#2#3{Nature (London)\ {\bf #1}  #2 (#3)}
\def\apjs#1#2#3{Astrophys.\ J.\ Supp.\ {\bf #1}  #2 (#3)}
\def\aipcp#1#2#3{AIP Conf.\ Proc.\ {\bf #1}  #2 (#3)}
\def\njp#1#2#3{New\ J.\ Phys.\ {\bf #1} (#3) #2}

\end{document}